\documentclass[useAMS,usenatbib,usegraphicx]{mn2e}
\usepackage{amsmath,aas_macros} % with this, one can use ADS bibtex entries in MNRAS

\voffset =- .6in %for Arxiv use

\title[Amplification by shocks in clusters]{Magnetic field amplification by shocks in galaxy clusters: application to radio relics}

\author[L. Iapichino and M. Br\"uggen]{Luigi Iapichino$^{1}$\thanks{E-mail: luigi@uni-heidelberg.de} and Marcus Br\"uggen$^{2}$\\
$^{1}$Zentrum f\"ur Astronomie der Universit\"at Heidelberg, 
Institut f\"ur Theoretische Astrophysik, Albert-Ueberle-Str.~2, D-69120 \\
Heidelberg, Germany\\
$^{2}$Jacobs University Bremen, Campus Ring 1, D-28759 Bremen, Germany}

\begin{document}

\date{Accepted 2012 April 11. Received 2012 April 11; in original form 2011 October 7}

\pagerange{\pageref{firstpage}--\pageref{lastpage}} \pubyear{2012}

\maketitle

\label{firstpage}

\begin{abstract}
Merger shocks induce turbulence in the intra-cluster medium (ICM), and, under some circumstances, accelerate electrons to relativistic velocities to form so-called radio relics. Relics are mostly found at the periphery of galaxy clusters and appear to have magnetic fields at the $\mu{\rm G}$ level. Here we investigate the possible origins of these magnetic fields. Turbulence produced by the shock itself cannot explain the magnitude of these fields. However, we argue that if the turbulent pressure support in the ICM upstream of the merger shock is of the order of 10 to 30 percent of the total pressure on scales of a few times $100\ {\rm kpc}$, then vorticity generated by compressive and baroclinic effects across the shock discontinuity can lead to a sufficient amplification of the magnetic field. Compressional amplification can explain the large polarisation of the radio emission more easily than dynamo turbulent amplification. Finally, clumping of the ICM is shown to have a negligible effect on magnetic field amplification.
\end{abstract}


\section{Introduction}
\label{intro}

In the classification of diffuse synchrotron radio emission in galaxy clusters, radio relics are polarised (generally at a level of $10$ to $30\%$) sources with a steep spectral index $\alpha \ga 1$. They have elongated shapes, with sizes between $10^2$ and $10^3\ {\rm kpc}$, and are usually located at the cluster periphery (\citealt{fgs08}, for a review). According to the scheme proposed by \citet{kbc04}, there are two classes of relics, the ``radio ghost'' caused by shock-induced compression of fossil radio plasma \citep{eg01,eb02} and the ``radio gischt'' caused by  shock accelerated cosmic ray (CR) electrons \citep{ebk98,hb07}.  In this paper, we will only be concerned with gischt relics. 

Relics are thought to be caused by shocks, triggered by a major merger and propagating through the intra-cluster medium (ICM). In a few cases (A521, \citealt{gvm08}; A3667, \citealt{fsn10}; A754, \citealt{mmg11}) the radio emission in relics is co-located with a shock front, detected by X-ray observations. 

Some observations suggest that relics host relatively large magnetic fields, with typical values of a few $\mu$G (for example, \citealt{bgf09, fsn10}). These fields are derived using several methods, mostly equipartition arguments, comparison of hard X-ray and radio emission (cf.~\citealt{fgs08}), the spectral aging \citep{mgb05} and the width of relics seen edge-on. The last method has been employed for a relic recently observed in the cluster CIZA J2242.8+5301 \citep{wrb10}. This relic is extreme in terms of its very regular morphology, the high degree of polarisation and the large magnetic field, between $5$ and $7\ \mu{\rm G}$.

The theoretical study of relics in cosmological simulations requires to connect the simulation output with a post-processing step for the computation of the radio emission. This exercise has been performed several times in the literature, using different hydrodynamical codes (mesh-based or SPH) and assumptions for the radio emission \citep{hby08,bps09,sho11}. Other related works focus on the acceleration mechanism of the CR electrons (a recent example is \citealt{kr11}), on predictions for future radio observations \citep{nhw11,vbw11} and on putting constraints on the merger geometry \citep{wbr11}.
The cited studies make assumptions for the magnitude of the magnetic field $B$ in the ICM, either based on a scaling of $B$ with density \citep{hby08,sho11}, or with thermal energy \citep{bps09}, because those simulations do not include any MHD treatment. However, there are a number of observations and simulations of magnetic fields in clusters, for the calibration of these scalings. Here we briefly review the results most relevant to our analysis.

Observations of relics show (cf.~\citealt{vrb09}) that the most extended, arc-like objects are found at large projected distances from the cluster centre. For example, CIZA J2242.8+5301 is a massive cluster ($L_{\rm X} = 6.8 \times 10^{44}\ {\rm erg\ s^{-1}}$, thence $T \simeq 9\ {\rm keV}$, \citealt{wrb10}), so its virial radius would be about $3\ {\rm Mpc}$, meaning that its relic is located approximately at $0.5\ R_{\rmn{vir}}$. Another prominent object, the northwest relic in A3667, has a projected distance of about $1\ R_{\rmn{vir}}$. These two cases set the distance range that this study investigates. The typical magnetic field near the lower edge of this range, derived by \citet{ckb01} from the statistics of Faraday rotation measure (RM) measurements, is $5\ \mu{\rm G}$ within a distance which, given the X-ray luminosities of the targets in their cluster sample, is approximately $0.3\ R_{\rmn{vir}}$. \citet{gmf06} study the RM in A2255 and, with their best model, derive a radial profile for $B$ with values between $0.6$ and $0.4\ \mu{\rm G}$ in the distance range between $0.5$ and $1\ R_{\rmn{vir}}$. Similarly, in their RM observations of Coma, \citet{bfm10} derive a profile for $B$ obeying $B(r) = B_0 [n(r)/n_0]^{\eta}$, where $n$ is the gas number density, and $B_0$ and $n_0$ are central values. Their best model is for $B_0 = 4.7\ \mu{\rm G}$ and $\eta = 0.5$, resulting in a value of $0.9\ \mu{\rm G}$ at $0.5\ R_{\rmn{vir}}$. With a different technique, based on de-polarisation arguments, \citet{bgf11} find in a sample of clusters on average $B \simeq 2.6\ \mu{\rm G}$ within approximately $0.3\ R_{\rmn{vir}}$.

In summary, observations support that the ambient magnetic field strength in the distance range relevant for radio relics is, within a factor of a few, $1\ \mu{\rm G}$ at  $0.5\ R_{\rmn{vir}}$.

These values are confirmed by cluster MHD simulations, performed with a variety of codes and prescription for additional physics \citep{dbl02, dt08,ds09,ddl09,xlc10,bds11,xlc11}. The theoretical results show a marked dependence on the magnetic seed, either primordial or produced by an AGN outflow, but under the constraint of a central magnetic field of a few $\mu{\rm G}$ they are generally in agreement with the observations.

In this work we want to address the following questions: How are the apparently large magnetic fields in relics produced? Can we explain the observed polarisation in relics? Our fiducial example will be the relic in the galaxy cluster CIZA J2242.8+5301?
 
An important process for magnetic field amplification, often invoked in the magnetogenesis of galaxy clusters, is the small-scale, turbulent dynamo \citep{dbl02,brs05,ssh06}.  However, we will show that the amplification of the magnetic field, $B$, is not driven by turbulence injected by the shock. Instead, we argue that ICM turbulence upstream of the shock produced a magnetic field, that is amplified by the shock. Thus the upstream magnetic field can be amplified by a factor close to the shock compression ratio or somewhat larger if baroclinic effects are important.

The role of shock-driven vorticity generation will be studied in Section \ref{jump}, where estimates for the downstream magnetic field and the polarisation level will be derived. The efficiency of turbulent amplification and the effect of clumping will be quantified in Section \ref{post-shock} and \ref{clumping}, respectively. 
The results will be discussed and summarised in Section \ref{discussion}.

\section{Vorticity jump across a propagating shock}
\label{jump}

\subsection{The vorticity equation and its terms}
\label{terms}

Hydrodynamical shocks can amplify vorticity $\bmath{\omega}$ which in the frozen-in case follows the same evolution equation as the magnetic field, $\bmath{B}$ (e.g.~\citealt{dw00}).
According to \citet{kp09} (their equation 5), the vorticity jump $\delta \bmath{\omega}$ across a propagating shock, along the binormal direction $\bmath{b}$ (tangential to the shock surface) is:
\begin{equation}
\begin{split}
\delta \bmath{\omega} \cdot \bmath{b} = \frac{\mu^2}{1+\mu}\frac{\partial M_{\rm S}}{\partial S}+\frac{1}{M_{\rm S}}\left ( \frac{\mu}{1+\mu}M_{\rm S}^2 -1 \right )\\
\times\left ( \frac{\partial \frac{1}{2} M_{\rm t}^2}{\partial S}+ \bmath{\omega}\times  \bmath{u} \cdot  \bmath{s} \right ) + \mu  \bmath{\omega} \cdot \bmath{b}
\end{split}
\label{vort-jump}
\end{equation}
where $\bmath{s}$ is tangential to the shock and $\bmath{n}$ is the shock-normal direction, so that the binormal direction is $\bmath{b} = \bmath{s} \times \bmath{n}$. We note that the normal component of the vorticity is continuous across a shock. The normalised density jump across the shock is $\mu = \rho_2/\rho_1 - 1$, where $\rho_2$ and $\rho_1$ are the pre-shock and post-shock gas densities, respectively, $M_{\rm t}$ is the turbulence Mach number of the upstream flow, $M_{\rm S}$ the Mach number of the shock and $\partial/\partial S$ is the spatial derivative in the direction tangential to the shock surface.

The first term on the right-hand side of equation (\ref{vort-jump}), the so-called curvature term, is relevant at locations where the shock speed varies along its surface, or where this surface bends.
The second term governs the baroclinic generation of vorticity, as it clearly appears in the derivation performed by \citet{k97}, but it is cast in a more general form than $\nabla \rho \times \nabla P$ (e.g., \citealt{rkc08}); in particular, the expression in equation (\ref{vort-jump}) is suited to deal with the shock propagation in a non-uniform flow. 
The third term is associated with flow compression at the shock, and arises from the conservation of angular momentum. We stress that this term in equation (\ref{vort-jump}) is non-vanishing only if the upstream fluid is not irrotational (i.e., if it has a non-vanishing vorticity); on the other hand, \citet{k97} shows that even a straight shock can generate vorticity in an irrotational fluid by baroclinic effects.

If the upstream flow is turbulent and the incoming shock is curved, all terms can potentially contribute to the amplification. The compressional contribution has the same nature as the magnetic amplification at shocks following from flux conservation, described by \citet{ebk98}. Consider a magnetic flux tube with strength $B_1$, oriented at an angle $\theta_1$ with respect to the shock normal: this tube will be bent by the interaction with the shock and the field amplified, according to 
\begin{equation}
\label{ebk01}
\tan \theta_2 = R \tan \theta_1
\end{equation}
and
\begin{equation}
\label{ebk02}
B_2 = B_1 \left( \cos^2 \theta_1 + R^2 \sin^2 \theta_1 \right)^{1/2}
\end{equation}
where the subscripts 1 and 2 refer to the pre-shock and the post-shock region, respectively, and $R = \rho_2 / \rho_1$ is the shock compression ratio. From equation (\ref{ebk02}) it is easy to see that the magnetic field amplification is negligible when $B_1$ is oriented parallel to the shock normal ($\theta_1 \sim 0$, thereby $B_2 \simeq B_1$), whereas $B_2 \simeq R B_1$ when $B_1$ is nearly parallel to the shock surface. The shock compression amplifies therefore preferentially the field along the shock surface, although only with a moderate efficiency namely, at most, $R = (\gamma +1) / (\gamma -1)$ for strong shocks. For a typical merger shock with Mach number $M_{\rm S} \sim 3$ the Rankine-Hugoniot shock jump relations predict $R \sim 3$, assuming a polytropic index $\gamma = 5/3$. Hence, shock compression results in an amplification of $B$ by a factor of a few. If the shock is further modified by cosmic rays, the compression factor can become even higher (cf.~Section \ref{discussion}). 

How strong is the vorticity produced by the curvature of the shock? According to \citet{t52}, the jump induced by shock curvature can be rewritten as
\begin{equation}
\label{truesdell}
\delta \omega = - \frac{\mu^2}{1+\mu} U_{\rm S} K
\end{equation}
where $U_{\rm S}$ is the fluid velocity tangential to the shock, in the shock reference frame, and $K$ is the shock curvature. For a simple estimate in the merger shock case, let $K \sim 1/D$, where $D$ is the distance of the shock from the cluster centre, set to $1.5\ {\rm Mpc}$ for similarity with CIZA J2242.8+5301. We also approximate $U_{\rm S}$ with  the turbulent velocity in the upstream medium. To be more quantitative, the ratio of turbulent to thermal pressure at the length scale $l$ can be written as

\begin{equation}
\label{pressure}
\frac{P_{\rmn{turb}}}{P_{\rmn{therm}}}(l) = \frac{v_{\rmn{turb}}^2(l) /
  3}{k T / (\mu m_{\rmn{p}})}  \,\,,
\end{equation}
where $v_{\rmn{turb}}(l)$ is a turbulent velocity at the scale $l$, $k$ is the Boltzmann constant, $\mu = 0.6$ is the mean molecular weight in a.m.u., $m_{\rmn{p}}$ is the proton mass, and $T$ is the gas temperature. 
Both simulations and observational evidence suggest that, in regions around the virial radius, $P_{\rmn{turb}}/P_{\rmn{therm}}\sim 0.1-0.3$ at the turbulence injection length scale $l_0$ of a few times $100\ {\rm kpc}$ (e.g., \citealt{vbg11}). In the following, the value $P_{\rmn{turb}}/P_{\rmn{therm}} = 0.2$ will be used, which corresponds to a turbulent velocity $v_{\rmn{turb}}(l_0) = 640\ {\rm km\ s^{-1}}$ for a typical ICM temperature of $5 \times 10^7\ {\rm K}$.

For merger shocks, we wish to study scales $l_{\rm s}$ smaller than the injection scale $l_0$. We set $l_{\rm s} / l_0 = 0.1$, so that $l_{\rm s}$ is of the order of a few tens ${\rm kpc}$, comparable to the relic width. The turbulent velocity $v_{\rm turb}(l_{\rm s})$ is derived from $v_{\rm turb}(l_0)$ by assuming Kolmogorov scaling, i.e.~$v(l) \propto l^{1/3}$. In this way,
\begin{equation}
\label{truesdell-ours}
\delta \omega \simeq \frac{\mu^2}{1+\mu} \frac{v_{\rm turb}(l)}{D} = \frac{\mu^2}{1+\mu} \frac{v_{\rm turb}(l_0)}{D} \left( \frac{l_{\rm s}}{l_0} \right)^{1/3}    \,\,.
\end{equation}
The value of $\delta \omega$, for $l_{\rm s}/l_0 = 0.1$ and $v_{\rm turb}(l_0)= 640\ {\rm km\ s^{-1}}$, is below $10^{-17}\ {\rm s^{-1}}$.
 For comparison, typical values of small-scale vorticity in turbulent regions of the ICM are about  $10^{-16}\ {\rm s^{-1}}$ \citep{krc07,pim11}. For this reason, we can neglect the injection of vorticity due to shock curvature. Locally, the curvature term can be significant at locations where the derivative $\partial M_{\rm S}/ \partial S$ is large, namely where $M_{\rm S}$ varies a lot along the shock surface. An example is given by  the large vorticity amplification at the interface between large-scale structure filaments and propagating merger shocks \citep{pim11}. Interestingly, bright notch-like features have also been observed at the edges of some radio relics \citep{rwh97,bdn06}.

\begin{figure}
  \resizebox{\hsize}{!}{\includegraphics{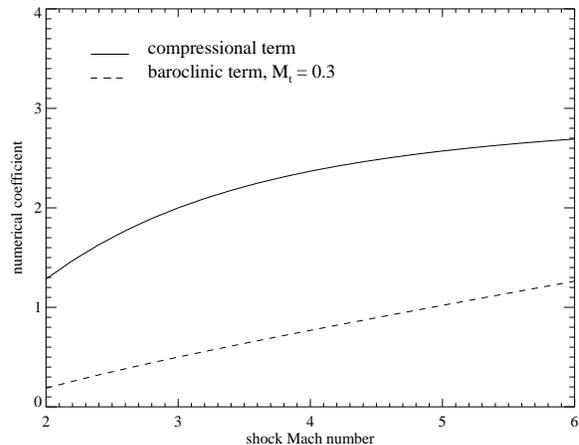}}
  \caption{Comparison of the numerical coefficients in front of the compressive and baroclinic terms (equation \ref{vort-jump}), described in the text, as a function of the shock Mach number $M_{\rm S}$. The baroclinic coefficient has been multiplied by the turbulence Mach number $M_{\rm t}$, whose value has been set to 0.3 (dashed line). The solid line refers to the compressive coefficient, namely the normalised density ratio $\mu$.}
  \label{jump-terms}
\end{figure}

As for the baroclinic term, a qualitative estimate of the relative weight of this term as a function of the shock Mach number $M_{\rm S}$, can be performed by a comparison of the numerical coefficients of the baroclinic and compressive terms in equation (\ref{vort-jump}), which are $1/M_{\rm S}\ (\mu/(1+\mu)M_{\rm S}^2 -1)$ and $\mu$, respectively.  Although a detailed analysis would require more information on the upstream flow and a complete computation of the terms in equation (\ref{vort-jump}), this task is beyond the scope of this paper.
For consistency, we further  multiply the numerical factor of the baroclinic term by $M_{\rm t}$, in order to roughly take into account the magnitude of $\bmath{u}$ in the product ${\bmath{\omega}} \times \bmath{u}$ in equation (\ref{vort-jump}), where the flow velocity $\bmath{u}$ is normalised to the sound speed. For the time being, $\partial(1/2\ M_{\rm t}^2) / \partial S$ in the baroclinic term is neglected. The normalised density ratio $\mu$ is expressed as a function of $M_{\rm S}$ according to the Rankine-Hugoniot relations (cf.~\citealt{kp09}):
\begin{equation}
\label{mu-ms}
\mu = \frac{M^2_{\rm S} -1}{1 + 1/2 (\gamma -1)M^2_{\rm S}}
\end{equation}
The range of expected turbulent pressure ratios in clusters corresponds to a range of $M_{\rm t}$ between 0.2 and 0.35  on a scale of a few times $10\ {\rm kpc}$. A value of 0.3 has been used for the comparison between the baroclinic and compressive coefficients of equation (\ref{vort-jump}), shown in Fig.~\ref{jump-terms}.

As expected, $\mu$ tends to 3 for increasing $M_{\rm S}$. The amplification of $\omega$ is mostly compressional. The baroclinic contribution is small, as expected for a subsonic turbulent flow (cf.~\citealt{db11}), but not negligible, especially at high values of $M_{\rm S}$: for $M_{\rm t} = 0.3$ and $M_{\rm S} = 5$, the baroclinic coefficient is 40\% of the compressional one. In this regime, both the compressional and baroclinic term contribute to the shock amplification of vorticity. 

In general, merger shocks are efficient in amplifying pre-existing vorticity, while they are less efficient in producing it via baroclinic effects. The estimate of the magnetic field in the post-shock region thus depends on the upstream field, which we will estimate in the next Section.

\subsection{Upstream turbulent flow in the outer ICM and resulting magnetic field}
\label{turbulent-icm}

As known from MHD cosmological simulations, under reasonable assumptions for the seed field, it is not possible to reproduce the magnetic field expected in clusters only by adiabatic compression caused by collapse during structure formation \citep{dgs05}. In this Section, with the use of simple analytical estimates, we demonstrate that the turbulent dynamo is able to boost the level of ICM magnetisation above the adiabatic compression values. These results are in rough agreement to the findings of MHD simulations (cf.~Introduction). 

Turbulence is driven by structure formation at large length scales (of the order of $l_0$, a few times $100\ {\rm kpc}$), and cascades to the dissipation length scale, presumably located at ${\rm kpc}$ or sub-${\rm kpc}$ scales. All length scales in this range contribute to the amplification of $B$, but in a Kolmogorov cascade (or any other similar scaling) most of the kinetic energy is contained at the largest scales, although the amplification is faster for smaller eddies (cf.~\citealt{ssh06}). In this simple analysis, we will restrict ourselves to the amplification by the largest scale eddies ($l = l_0$).

The maximum magnetic field attainable by the turbulent dynamo is some fraction $f$ of the equipartition value with the kinetic energy (see \citealt{s98, ssh06, sbs10, fcs11}): 
\begin{equation}
\label{saturated}
B_{\rm sat} = \sqrt{f\, 8 \pi\, \rho\, e_{\rm kin}} = \sqrt{f\, 4 \pi\, \rho\, v^2}
\end{equation}
In \citet{isn11} it is shown that the stirring in the cosmic gas is driven by a combination of compressional and solenoidal modes, with the latter ones dominating in the ICM. For subsonic, solenoidally driven turbulence \citet{fcs11} derived $f \simeq 0.4$; we will use this value in equation (\ref{saturated}). 

Furthermore, we note that turbulence in clusters is expected not to be volume-filling. According to \citet{ssh06}, for a volume-filling factor $f_{\rmn{V}}$ the rms magnetic field is equal to 

\begin{equation}
\label{saturated-ii}
B_{\rm sat, V} = B_{\rm sat} \times f_{\rmn{V}}^{1/2} = \sqrt{f_{\rmn{V}}\, f\, 4 \pi\, \rho\, v^2}
\end{equation}
Here we set $f_{\rmn{V}} = 0.3$ \citep{in08}. 

We use equation (\ref{saturated-ii}) for obtaining an estimate of the magnetic field in the ICM, before the shock propagation. We set $v_{\rmn{turb}}(l_0) = 640\ {\rm km\ s^{-1}}$, as in Section \ref{terms}. As for the density in equation (\ref{saturated-ii}), the two values $\rho = 10^{-27}$ and $10^{-28}\ \rmn{g\ cm^{-3}}$ will be taken as representatives of the density at central distances $r = 0.5\ R_{\rmn{vir}}$ and $r = R_{\rmn{vir}}$, respectively. The results are values of the saturation magnetic field $B_{\rm sat, V} = 2.5\ \mu{\rm G}$  at $r = 0.5\ R_{\rmn{vir}}$,  and $0.8\ \mu{\rm G}$ at $r = 1\ R_{\rmn{vir}}$.  

These values are about a factor of two larger than the typical results from observations and MHD simulations, but our simple derivation cannot represent faithfully the whole complexity of the turbulent dynamo. Firstly, it is not guaranteed that the field is saturated up to the scale $l_0$, and this could slightly reduce our estimates. Furthermore, especially in the cluster periphery, a larger share of compressionally-driven turbulence might result in a smaller value for $f$ in equation (\ref{saturated-ii}).

\subsection{Magnetic field and polarisation in the post-shock region}
\label{bpol-psr}

Having shown that the compressional term is dominant for vorticity production across the shock, we go back to the question of the amplification of $B$, drawing some estimates from equations (\ref{ebk01}) and (\ref{ebk02}). Assuming that the upstream flow is isotropic, it is easy to see that the two $B$ components parallel to the shock surface are amplified by a factor $R$, while the normal component is not amplified. The strength of the downstream field is thus

\begin{equation}
\label{estimate-ampl}
B_2 =  \left ( \frac{2 R^2 B_1^2}{3} + \frac{B_1^2}{3} \right )^{1/2} = B_1 \sqrt{\frac{2 R^2 + 1}{3}}  \,\,.
\end{equation}

For $R = 3$, the shock amplification predicted by equation (\ref{estimate-ampl}) is about 2.5, without taking into account some smaller contribution by baroclinicity. Given this amplification and the upstream estimates for $B$ at the end of Section \ref{turbulent-icm}, $B$ reaches a post-shock value of $2.0\ \mu{\rm G}$ at $r = R_{\rmn{vir}}$, and of $6.2\ \mu{\rm G}$ at $r = 0.5\ R_{\rmn{vir}}$.  For these estimates the same caveat mentioned at the end of Section \ref{turbulent-icm} applies. We defer a more detailed comment to Section \ref{discussion}.

Another effect of the compressional amplification is to make the downstream magnetic field anisotropic, by acting preferentially on the field component parallel to the shock. The degree of order introduced in the downstream field can be related to the polarisation observed in radio relics. According to \citet{bgf09} (see also \citealt{b66}), in this case the observed polarisation is 

\begin{equation}
\label{pol}
P_{\rmn{obs}} = \frac{3 \delta + 3}{3 \delta + 7} \frac{1}{1 + (B^2_{\rmn{r}} /B^2_{\rmn{o}})}    \,\,,
\end{equation}
where $\delta = 2 \alpha +1$, $\alpha$ being the relic spectral index. The random and ordered components of the downstream magnetic fields are $B_{\rmn{r}}$ and $B_{\rmn{o}}$, respectively. In the following, we label the components of $B$ parallel to the shock surface as ordered, and the normal component as random; from equation (\ref{estimate-ampl}), the ratio $B^2_{\rmn{r}} /B^2_{\rmn{o}}$ is set therefore to $1/(2R^2)$. The first factor of the right-hand side of equation (\ref{pol}) can be interpreted as the intrinsic relic polarisation; for a typical value of $\alpha = 1$, it is equal to 0.75. The second factor of the right-hand side is always smaller than unity and represents a suppression of the intrinsic polarisation. For our fiducial choice $R = 3$, this factor is equal to 0.95, and $P_{\rm obs} = 0.71$. 

In this oversimplified estimate we neglected to introduce the relic viewing angle, which further suppresses $P_{\rm obs}$ \citep{ebk98}. For the case of relics seen almost edge-on on the plane of sky, this suppression is not dominant, and in fact the polarisation derived from equation (\ref{pol}) is not far from that measured in CIZA J2242.8+5301 ($50$--$60\%$ level, \citealt{wrb10}). In relics with a less favourable viewing geometry, or more complicated morphology, the polarisation is lower (10 to $30\%$).

Despite of the large uncertainties of the estimate in equation (\ref{pol}), our aim here was to show that the ordering introduced by compressional amplification has the potential of producing a large polarisation of the radio emission, at odds with turbulent magnetic amplification, which acts in the opposite direction of depolarising the signal, as observed in radio halos (\citealt{mgf04,bgf11}).

Finally, we note that the possibility of a turbulent and magnetised pre-shock medium is not in contradiction with the regular morphology of many radio relics. Hydrodynamical simulations of shock-turbulence interactions (e.g., \citealt{llm93}) show that the shock morphology is mainly unaltered, as long as $M_{\rm S}$ is sufficiently large in comparison to $M_{\rm t}$.

\section{Turbulent amplification in the post-shock region}
\label{post-shock}

Here we show that, if the upstream flow is not turbulent, the stirring induced by the propagation of a single merger shock is not able to produce fully developed turbulence in the downstream region, and hence turbulent amplification is not efficient. 

In hydrodynamical simulations of forced turbulence (e.g., \citealt{sfh09,fks09}), turbulence is fully developed after a few large-scale eddy turnover times $t_{\rm eddy}$, given by $t_{\rm eddy} = L / v$, where $L$ is the integral length (the largest length scale at which the flow is stirred) and $v$ is the typical velocity at the scale $L$. 

During the propagation of the merger shock through the cluster outskirts, the downstream region is stirred. Let $t_{\rm evol}$ be defined as the typical evolutionary timescale of the post-shock region. Turbulence is considered to be fully developed if the ratio 
\begin{equation}
\label{number}
N = \frac{t_{\rm evol}}{t_{\rm eddy}}
\end{equation}
is larger than a few. 
Here, we interpret $t_{\rm evol}$ as the advection timescale of the downstream gas over the width of the radio relic, $L_{\rm relic}$, or the region where a substantial amplification of the magnetic field is required. Thus we write 
\begin{equation}
\label{evolutionary}
N = \frac{L_{\rm relic}}{L_{\rm eddy}}  \frac{v_{\rm turb}(L_{\rm eddy})}{v_{\rm down}}
\end{equation}
where $v_{\rm down}$ is the downstream velocity in the shock reference frame.  For shocks with Mach numbers between 2 and 5 and shock velocities $v_{\rm S} \sim 1000 -  4000\ {\rm km\ s^{-1}}$, $v_{\rm down} = 1000\ {\rm km\ s^{-1}}$ is a reasonable value. 
Let us assume a maximal relic width of $200\ {\rm kpc}$ and a Kolmogorov scaling, i.e. $v(l) \propto l^{1/3}$ with a turbulent velocity of $v_{\rm turb}(200\ {\rm kpc}) = 100\ {\rm km\ s^{-1}}$ which is motivated by simulations of  shock propagation through a laminar upstream region \citep{pim11}\footnote{The adaptive mesh refinement in \citet{pim11} is designed to effectively refine the post-shock region, but it is not equally effective upstream. For this reason, here we consider the downstream turbulent velocity reported by \citet{pim11} as representative of the stirring produced by a shock propagating in a nearly laminar upstream flow. This is exactly the case that we want to study in this Section, although in real cluster environments the turbulent velocity is expected to be much higher.}. 
The turbulent velocity that we need to consider here is the turbulent velocity that results from a largely laminar inflow. Hence, we find that turbulence is not fully developed, i.e.~$N<2$ as long as $L_{\rm eddy} > 2\ {\rm kpc}$. In cluster outskirts, the dissipation scale is very uncertain, but expected to be close to the kiloparsec scale \citep{snb03}.

The (unsurprising) result of this analysis is that the flow in the post-shock region, resulting from a laminar flow upstream, does {\it not} show fully developed turbulence, mainly because the timescale $t_{\rm evol}$ is too short.  Clearly, if the upstream flow itself is turbulent, mere compressional amplification may suffice as shown in  Section \ref{terms}. As a side note, the compressional driving of turbulence, provided by a propagating shock, is rather inefficient in amplifying $B$ \citep{fcs11,jpr11}.

\section{Density inhomogeneities in the upstream medium}
\label{clumping}

Based on {\it Suzaku} observations of the regions close to the virial radius, it has been suggested that there is some level of clumping in the outer ICM \citep{uws11, sam11}. Numerical simulations by \citet{nl11} have shown that clumping in the ICM, as for the substructure resolved in that work, starts being relevant at $r \simeq R_{200}$, and thus affects mostly the outer part of the distance range studied throughout this paper. Gas clumps in the ICM have recently been invoked by \citet{bd11} in their mechanism for balancing radiative cooling in cool-core clusters. Moreover, they estimated that about 5 per cent of the accreted baryons are in gas clumps with masses of $\sim 10^8\ {\rm M_{\odot}}$.

As well as for the turbulent medium, the propagation of a shock through an inhomogeneous medium can be a source of additional vorticity and thus magnetic amplification. Vorticity is generated by the baroclinic mechanism at density inhomogeneities \citep{sz94}. In \citet{iyi09}, for conditions typical of supernova remnants, the interaction between a strong shock wave and a multiphase interstellar medium produces amplifications well above the level attainable by turbulence.

Here, we make a few geometrical estimates: a sphere with $R = 1\ {\rm kpc}$ and density $\rho_{\rm b} = \chi \bar{\rho}$, where the density contrast $\chi$ is set to 10 and $\bar{\rho}$ is the ambient density in the outer ICM (set to $10^{-28}\ {\rm g\ cm^{-3}}$, i.e.~about 200 times the average baryon density), has a mass $M_{\rm b} = 6 \times 10^4\ M_{\odot}$. As a working hypothesis, we assume that 5\% of the baryons in the outer ICM is in such clouds. We consider a volume of $(0.05 \times 1.7 \times 1.7)\ {\rm Mpc}$, similar to the volume of the CIZA J2242.8+5301 relic, with an average density $\bar{\rho}$. Some simple algebra shows that in this volume there are $1.8 \times 10^5$ such clumps. From geometrical considerations, their average spacing will be around $10\ {\rm kpc}$. According to \citet{pfb02}, the ratio between the clump radius and this typical separation is smaller than the minimum value needed for the clumps to interact and merge in their late evolution. The initial volume fraction occupied by the clumps is $V_{\rm b,in} = 0.005$. By visual inspection of the results of \citet{pfb02}, it is assumed that the clump gets spread in a volume $V_{\rm b,final} = 10 V_{\rm b,in}$ before being dispersed. This results in a final volume filling factor which remains small (below the 10\% level).

By flux conservation, we set the initial field in the clumps as $B_{\rm b,in} = B_{\rm up} (\rho_{\rm b} / \bar{\rho})^{2/3} \sim 3 B_{\rm up}$, where $B_{\rm up}$ is the upstream magnetic field in the ICM.

In their MHD simulations of shock-cloud interaction, \citet{sss08} find that, after the shock propagation and before the complete clump dispersal, the clump magnetic field is amplified up to a value $B_{\rm b,final} = f_{\rm b} B_{\rm b,in}$, with the amplification factor $f_{\rm b}$ which depends on the orientation of $B$ and on the $\beta$ parameter, defined as $\beta = P / P_{B}$, the ratio between the thermodynamical and magnetic pressure. For the problem parameters considered here,
\begin{equation}
\label{beta}
\beta = \frac{P}{P_{B}} = \frac{8 \pi\, \rho_{\rm b}\, k T_{\rm b}}{\mu\, m_{\rm p}\, B_{\rm b,in}^2} \simeq 8 
\end{equation}
where we used $T_{\rm b} = 5 \times 10^6\ {\rm K}$ (smaller than the ambient $T$ by a factor of $\chi$ for ensuring pressure equilibrium between the ICM and the clumps), and $B_{\rm b,in} = 1.5\ \mu{\rm G}$, resulting from $B_{\rm up} = 0.5\ \mu{\rm G}$. The value of $\beta$ derived in equation (\ref{beta}) is within the range examined by \citet{sss08}; the corresponding amplification factor is set to $f_{\rm b} = 2$ (in line with the cited results, but somewhat extreme).

According to \citet{sss08} and many other similar studies, the typical timescale for the clump disruption (and for the $B$ amplification) is set by the so-called cloud-crushing time, which is equal to clump shock crossing timescale times $\chi^{1/2}$. Given the large difference in size between the clump and the post-shock region, it is obvious that the dispersal mechanism (and the related $B$ amplification) has enough time to develop during the shock propagation in the post-shock region.  

Summarising the arguments provided above, we arrange them in an estimate of the $B$ amplification. First, we set $B_{\rm down} = 2.5 B_{\rm up}$ as the downstream value of amplified the magnetic field, without the effect of clumping (cf.~equation \ref{estimate-ampl}). The volume-weighted downstream $B$, including clumping, is $(1 - V_{\rm b,final}) B_{\rm down} + V_{\rm b,final} (B_{\rm b,final}) = 1.07\ B_{\rm down}$. The whole effect of the clumping considered here is an amplification at the negligible level of some percent. 

There are many arbitrary parameters in this analysis, but the amplification level will not be significantly larger without invoking a drastically higher baryon fraction in these clumps, or larger sizes. The latter is unlikely as it would destroy the smooth morphology of the radio relic \citep{wbr11}.

\section{Discussion and conclusions}
\label{discussion}

In this work, we examined the amplification of magnetic fields in the post-shock region of merger shock waves in galaxy clusters. We considered several mechanisms for amplifying magnetic fields in the downstream region of the shock, including small-scale dynamo action, compressional amplification, baroclinic amplification and amplification by the interaction with density inhomogeneities.

We demonstrated that turbulent amplification is inefficient in producing magnetic fields of the magnitude required by observations in radio relics because the stirring induced by the shock propagation in the post-shock region is not sufficient to drive fully developed turbulence. Instead, we propose that the shock propagates through regions of the outer ICM with a turbulent pressure support of the order of 10 to 30 percent.

The dominant mechanism is the compressional amplification, which amplifies the magnetic field components parallel to the shock surface. 
Together with compression at shocks, baroclinic effects are expected to be marginally relevant for high Mach numbers ($M_{\rm S} \ga 3$) and intense ICM turbulence ($M_{\rm t} = 0.3$).
Motivated by recent observational indications, we explored in Section \ref{clumping} the role of clumping for the magnetic field amplification, which turns out to be modest. 

Compressional amplification can suffice to explain magnetic fields in radio relics.
Upstream fields are amplified by the outgoing merger shock, as expressed in Section \ref{jump} by equation (\ref{estimate-ampl}). For a typical shock Mach number $M_{\rmn S} = 3$, the resulting amplification factor is equal to 2.5. The magnetic field in the post-shock region is thus estimated to be $6.2\ \mu{\rm G}$ at $r = 0.5\ R_{\rmn{vir}}$, and $2.0\ \mu{\rm G}$ at $r = 1\ R_{\rmn{vir}}$.

These values are in good agreement with the available estimates of magnetic fields in radio relics (e.g., \citealt{bgf09,whr11}). In the case of CIZA J2242.8+5301, the field from observations is between 5 and $7\ \mu{\rm G}$, which is well matched by our prediction. 
 There is some discrepancy between observations and theory for the NW relic of A3667, where the observational estimate for the relic field is $B = 3\ \mu{\rm G}$ at  $r \sim 1\ R_{\rmn{vir}}$ \citep{fsn10}, to be compared to a theoretical prediction at the level of about $2\ \mu{\rm G}$. Given the simplicity of our model, this still constitutes reasonable agreement.

The proposed model for the compressional amplification of the magnetic field, in contrast with turbulent dynamo models, can naturally account for a large polarisation of the radio emission (equation \ref{pol}), although more realistic estimates should incorporate the role of the relic viewing angle \citep{ebk98}.

Compared to X-ray observations, the outer regions of clusters can potentially be better studied by Sunyaev-Zel'dovich (SZ) observations because the SZ effect depends only on the electron density to the first power, while the X-ray emission depends on the electron density squared. Facilities such as the {\it Atacama Cosmology Telescope} ({\it ACT}), the {\it South Pole Telescope} ({\it SPT}) and the {\it Planck} satellite are searching for the SZ signal of galaxy clusters and have found some interesting first results (see, e.g., \citealt{bsw11}). For example, it was found that the SZ signal caused by clusters is by a factor of $\sim 2$ smaller than predicted by models of clusters that suggest that the pressure by the electrons has been overpredicted \citep{lrs10}. This is presumably caused by a substantial nonthermal pressure at cluster outskirts, most of it is likely to be turbulent pressure (see e.g.~\citealt{snb10}). This picture is also supported by simulations \citep{lkn09,vbk09,pim11}. However, the volume-filling factor of turbulence in the cluster outskirts is still debated \citep{valda11,isn11}.

Currently, there are few measurements of magnetic fields in regions so far from the cluster centre,  such as for example \citet{ckb01}, who detected some Faraday RM excess out to central distances around $0.5\ {\rm Mpc}\ h^{-1}$. However, with the advent of the {\it SKA}, the magnetic fields in such regions can be better probed, by measuring the Faraday rotation in a much larger number of background sources \citep{kab09}. Synchrotron radiation propagating through a magnetised plasma undergoes Faraday rotation, and the rotation measure RM is given by:
\begin{equation}
\label{rm}
{\rm RM} = 812 \int _0 ^{L_{\rm path}} n_{\rm e} B_{\parallel} dl\ {\rm rad\ m^{-2}}
\end{equation}
where $L_{\rm path}$ is the path along the line of sight in units of ${\rm kpc}$, $n_{\rm e} = \rho / (m_{\rm p} \mu_{\rm e})$ is the electron density in ${\rm cm^{-3}}$ ($m_{\rm p}$ is the proton mass, $\mu_{\rm e} \simeq 1.14$ is the average mass per electron in a.m.u), and $B_{\parallel}$ is the component of $B$ along the line of sight, in $\mu{\rm G}$. We assume that a tangled magnetic field topology in the ICM leads to a Gaussian distribution of RMs with a vanishing mean $\langle {\rm RM} \rangle$ and a RM variance $\sigma_{\rm RM}^2$ given by \citep{mgf04}:
\begin{equation}
\label{rmvar}
\sigma_{\rm RM}^2 = \langle {\rm RM}^2 \rangle = 812^2 \Lambda \int (n_{\rm e} B_{\parallel})^2 dl
\end{equation}
where $\Lambda$ is the typical size of a turbulent cell. Under the assumption that the gas density profile follows a $\beta$-model
\begin{equation}
\label{betaprof}
n_{\rm e} = n_0 \left ( 1 + r^2 / r_{\rm c}^2 \right )^{3 \beta /2}
\end{equation}
with central density $n_0$ and core radius $r_{\rm c}$, the RM dispersion  $\sigma_{\rm RM}$ can be expressed as (\citealt{f96}; see also \citealt{dsg01}): 
\begin{equation}
\label{rmdisp}
\sigma_{\rm RM} = \frac{K B n_0 r_{\rm c}^{1/2} \Lambda^{1/2}}{\left ( 1 + r^2 / r_{\rm c}^2 \right )^{(6 \beta -1)/4}} \sqrt{\frac{\Gamma(3\beta - 0.5)}{\Gamma(3\beta)}} \,\,,
\end{equation}
where $\Gamma$ is the Gamma function, and $K$ is a numerical factor which depends on the location of the radio source with respect to the cluster, and is equal to 624 or 441 for background or embedded sources, respectively. The magnetic field $B$ is assumed to be constant in the ICM, although a radial dependence of the type $B(r) = B_0 [n(r)/n_0]^{\eta}$ \citep{bfm10} can be introduced by substituting $\beta$ with $\beta(1+\eta)$ in equation (\ref{rmdisp}) \citep{dsg01}.

We adopt the same  $\beta$-model parameters as in \citet{wbr11}: $r_{\rm c} = 134\ \rmn{kpc}$, $\beta = 2/3$, $n_0 = 2 \times 10^{-3}\ {\rm cm^{-3}}$, and we set $r = 1.5\ {\rm Mpc}$. The typical size of the turbulent eddies $\Lambda$ is set to $100\ \rm{kpc}$, according to \citet{ssh06} and to the assumptions of this work. Finally, we consider a background source ($K = 624$). The resulting RM dispersion for the assumed upstream field of $B = 2.5\ \mu{\rm G}$ is $\sigma_{\rm RM} = 9$, and for a shock-amplified field of $B = 6.2\ \mu{\rm G}$, it is $\sigma_{\rm RM} = 22$.

This estimate is probably too rough because the magnetic field has the amplified value only in the post-shock region, and not everywhere in the ICM along the line of sight. A further reduction of $\sigma_{\rm RM}$ comes from imposing a radial profile for $B$: for a line of sight at $r = 0.5\ R_{\rm vir}$ from the cluster centre, it results in $\sigma_{\rm RM}$ about 20\% smaller than in equation (\ref{rmdisp}), as estimated from applying it on Coma, using parameters from \citealt{bfm10}. On the other hand, for a line of sight passing through a radio relic, the assumption of a Gaussian distribution for the RM might not hold, because of the ordering introduced on $B$ by the compressional amplification. As a result, $\langle {\rm RM} \rangle$ could be non-vanishing for lines of sight crossing the relic.

The problem of Faraday rotation in connection to cluster merger shocks is interesting and worth being further studied, also by means of numerical simulations, because the derived values of $\sigma_{\rm RM}$, despite of uncertainties in their derivation, are within the sensitivity goal of upcoming instruments such as {\it SKA} or {\it ASKAP} \citep{bg04,kab09}.

Besides the mechanisms considered here, there are a number of other processes that are able to amplify the magnetic field at shocks. Viable channels are the Weibel instability \citep{w59} and the amplification driven by cosmic-ray acceleration (the so-called Bell instability; \citealt{lb00,b04}). The main issue with these mechanisms is, as in the case of clumping, the length scale: this is a known concern for the Weibel instability \citep{msk06}, which has the fastest growing mode below the parsec scale by orders of magnitude, although it has been speculated about a rapid transfer of magnetic energy to cosmological scales \citep{mff05}. As for the Bell instability, in the non-linear stage the dominant length scale can grow up to the CR Larmor radius \citep{rs09}. For the typical conditions considered in this work, we argue that this length scale remains below or around the parsec scale (cf.~equation 6 of \citealt{rs09}, and the discussion therein). 

Finally, \citet{bjl09} studied small-scale dynamos in the context of diffusive shock acceleration. They consider the effect of the CR shock precursor on the upstream density perturbations, and identify a mechanism that is able to generate an efficient amplification of the magnetic field in the precursor. In this model, an important condition is the existence of density perturbations in the upstream flow. In the framework of ISM studies, this is guaranteed by the relatively large Mach number of the flow. However, in subsonic flow, as in the ICM, the amplitude of the density perturbations is much smaller (see for example \citealt{fks08}), and the amplification in the precursor inefficient. For length scales and separations, the clumping described in Section \ref{clumping} cannot contribute to this mechanism. As a remedy, \citet{bjl09} hint at the possibility that density fluctuations are produced in the CR precursor itself. This needs to be studied in future work.

Faraday rotation measurements in the outer ICM will help to discriminate between purely hydrodynamical models of field amplification, which require some level of turbulence and magnetisation in the cluster outskirts, and other models, which link magnetic field amplification to CR acceleration.

\section*{acknowledgements}
MB acknowledges support from the Deutsche Forschungsgemeinschaft from the grant FOR1254. We are grateful to the referee, Klaus Dolag, for his insightful and constructive suggestions, which improved the presentation of this work. Thanks to V.~Antonuccio-Delogu and J.~C.~Niemeyer for having supported the first phases of this work, to A.~Bonanno and W.~Schmidt for useful comments on the manuscript, and to M.~Bartelmann, C.~Federrath, R.~Klessen and F.~Vazza for interesting discussions.

\bibliography{cluster-index}
\bibliographystyle{bibtex/mn-web}

\bsp

\label{lastpage}

\end{document}